\documentclass[a4paper,12pt]{article}

\usepackage{graphics}

\begin{document}

\title{
\begin{flushright}
{\normalsize IRB-TH-8/03}
\end{flushright}
\vspace{2 cm}
\bf Phantom appearance of non-phantom matter}

\author{ H. \v Stefan\v ci\'c\thanks{shrvoje@thphys.irb.hr}
}

\vspace{3 cm}
\date{
\centering
Theoretical Physics Division, Rudjer Bo\v{s}kovi\'{c} Institute, \\
   P.O.Box 180, HR-10002 Zagreb, Croatia}


\maketitle

\abstract{Two cosmological models with non-phantom matter having the same
expansion of the universe as phantom cosmologies are constructed. The first
model is characterized by the evolving gravitational ``constant" $G$ and a dark
energy component with a non-conserved energy-momentum tensor. The second model
includes two interacting components, the dark energy component and the matter
component. Closed form solutions are obtained for the constant values of model
parameters and constraints on the parameters of each model from
cosmological observations are outlined. For both models it is explicitly shown 
how the components of each model produce the expansion of the universe 
characteristic of phantom cosmologies, despite the absence of phantom energy. 
These
findings stress the interpretation of phantom energy as an effective description
of the more complex dynamics of non-phantom matter.}

\vspace{2cm}

\section{Introduction}

\label{sec1}

Complementary cosmological observations of supernovae of the type Ia (SNIa)
\cite{SN}, cosmic microwave background radiation (CMBR) \cite{CMBR}, large-scale
structure (LSS) \cite{LSS} and other cosmic phenomena have firmly established the 
picture of the accelerated expansion of the contemporary universe \cite{Rev}. 
The present accelerating phase of the expansion of the universe and its onset 
at a relatively low redshift ($z \sim 1$) represent one of the most intriguing
and most studied problems in modern cosmology. The majority of theoretical
explanations of this phenomenon invoke a new component of the universe named
{\em dark energy} \footnote{There are alternative explanations rooted in 
brane-world models which do not require dark energy \cite{singh}.} 
with the equation of state (EOS)  
\begin{equation}
\label{eq:eos}
p_{d} = w \rho_{d} \, ,
\end{equation}
where $\rho_{d}$ and $p_{d}$ represent dark energy density and pressure,
respectively. The most studied theoretical candidates for the role of dark
energy are the cosmological constant (CC) ($w=-1$) \cite{wein,peeb,pad}, and 
its dynamical variants such as the renormalization group running CC 
\cite{mi, sola, bonanno}, quintessence ($w \ge -1$) \cite{Q}, 
tachyon models \cite{tach} ($w \ge -1$) and the Chaplygin gas ($w \ge -1$) 
\cite{CG}. 

Recent analyses of cosmological observations 
\cite{odman, padman1, padman2,lima, sahni}
allow, and even favour, a sort of dark energy with a supernegative EOS, i.e.
$w < -1$. This unorthodox type of dark energy, first introduced in \cite{cald},
was named {\em phantom energy}. Many models of phantom energy appeared soon
\cite{tech}, addressing both its fundamental implications and cosmological
consequences. One of the most interesting features of phantom energy is
certainly the possibility of the divergence of the scale factor of the universe in
finite time. The expansion of the universe in such a model of phantom energy 
leads to the unbounding of all bound structures, a phenomenon also vividly 
referred to as ``Big Rip" \cite{cald2}. 

Although phantom energy represents a phenomenologically appealing possibility, 
the violation of the dominant energy condition (DEC), inherent in phantom
energy models, leads to problems at the microscopic level. For example, 
it is possible to describe phantom energy in terms of the effective scalar field 
theory with negative kinetic terms, valid up to some cut-off scale. In such a
formulation the vacuum of the theory is no longer stable, i.e. phantom energy
decays. Such theories can still be cosmologically viable if the lifetime of
phantom energy surpasses the age of the universe. This requirement puts 
stringent constraints on the parameters of the effective scalar filed theory,
above all on its cut-off scale \cite{dec1,dec2}. There still remains a question
whether some other viable microscopical formulation exists.

In such a conflict between favour from the observational side and disfavour from the
theoretical side, phantom energy models face an interesting alternative:
{\em possibility that matter which has no phantom characteristics (e.g. satisfies
DEC) produces observational effects attributed to phantom energy}. In this paper
we consider two realizations of this possibility. The first realization 
given in section \ref{sec2} is
a model reminiscent of {\em generalized phantom energy} \cite{GPE}, in which we
consider a sort of
cosmology with a time-dependent gravitational ``constant" $G$ and a
dark energy component with a non-conserved energy-momentum tensor. The second
realization is based on the dynamics of two interacting cosmological components
and is displayed in section \ref{sec3}.

\section{A model with an evolving G}

\label{sec2}

We consider a cosmological model with two components. The first component, 
which we call the matter component, has the equation of state
\begin{equation}
\label{eq:eosmatter}
p_{m}=\gamma(a) \rho_{m} \, ,
\end{equation} 
where $\gamma(a) \ge 0$, and $\rho_{m}$ and $p_{m}$ denote the energy density and
pressure of the first component, respectively. We assume that the
energy-momentum tensor of this component is conserved, 
$T^{\mu \nu}_{m ; \nu} = 0$, which leads to a well-known relation for the scaling
of $\rho_{m}$ with the scale factor $a$ \footnote{The subscript $0$ denotes the
present epoch throughout the paper.}:
\begin{equation}
\label{eq:rhomatter}
\rho_{m} = \rho_{m,0} \; e^{-3 \int_{a_{0}}^{a} (1+\gamma(a')) \frac{da'}{a'}}
    \, .
\end{equation}
The second component, which we call the dark energy component, satisfies DEC and
has the equation of state
\begin{equation}
\label{eq:eosdark}
p_{d}=\eta(a) \rho_{d} \, , \; \; \;\;\;\; \eta(a) \ge -1 \, .
\end{equation}        
Here $\rho_{d}$ stands for the energy density and $p_{d}$ denotes 
the pressure of the 
dark energy component. We assume that the energy-momentum tensor of the dark
energy component is not conserved, i.e. $T^{\mu \nu}_{d ; \nu} \neq 0$.
Therefore, the parameter of EOS (\ref{eq:eosdark}) does not determine the
scaling of $\rho_{d}$ with $a$. One possible way of harmonizing non-conservation
of $T^{\mu \nu}_{d}$ with the general covariance of the Einstein equation is
the promotion of 
the gravitational constant $G$ into a time-dependent function $G(t)$
(see reference \cite{GPE} for details) \footnote{Many models consider the
time-dependent $G$, such as the renormalization group running of $G$
\cite{mi,sola,bonanno,odin}, the time-dependence of $G$ originating from extra
dimensions \cite{extra} or the effective $G$ in scalar-tensor theories
\cite{scalten}.}.  
$G(t)$ satisfies a generalized conservation condition
\begin{equation}
\label{eq:gencov}
(G(t) T^{\mu \nu})_{;\nu} = 0 \, ,
\end{equation}
where $T^{\mu \nu} = T^{\mu \nu}_{m} + T^{\mu \nu}_{d}$. It is important to
stress that the procedure explained above does not represent some trivial 
multiplication of the constant
$G$ by some function of time $f(t)$ and multiplication of the total 
energy-momentum tensor $T^{\mu \nu}$
by $f(t)^{-1}$ since the energy-momentum tensor of the matter component is
conserved. The relation (\ref{eq:gencov}) can be expressed as
\begin{equation}
\label{eq:Geq}
d(G \rho_{d}) + \rho_{m} dG +3 G \rho_{d} (1+\eta(a))\frac{da}{a}  = 0 \, .
\end{equation}
This equation determines the dynamics of $G$ in terms of energy densities and
parameters of EOS of the components of the universe. At this place, it is important 
to notice that in the Friedmann equation
\begin{equation}
\label{eq:Friedmann}
\left(\frac{\dot{a}}{a} \right)^{2} + \frac{k}{a^{2}} = \frac{8 \pi}{3}
G (\rho_{m}+\rho_{d})\, ,
\end{equation}
the evolutions of both $G$ and energy densities $\rho_{m}$ and $\rho_{d}$
determine the kinematics of the universe, i.e. the function $a(t)$. 
The aim of this
section is to investigate the possibility that the product
$G(\rho_{d}+\rho_{m})$ has a component which grows with the scale factor (i.e.
its effective parameter of EOS is smaller than $-1$), while the components of the
universe have non-phantom nature (they satisfy DEC). 

To this end, we introduce an assumption of the following scaling behaviour:
\begin{equation}
\label{eq:Glaw}
G \rho_{d} = G_{0} \rho_{d,0} \left( \frac{a}{a_{0}} \right)^{-3(1+w(a))} \, ,
\end{equation}
where $w(a) < -1$. This assumption clearly introduces a source into the
Friedmann equation (\ref{eq:Friedmann}) which is identical with the source
originating from phantom energy with the parameter of EOS $w(a)$ in a
model with constant $G$.  

The evolution equation for $G$ becomes
\begin{eqnarray}
\label{eq:Glaw2}
dG & = & \frac{3 G_{0} \rho_{d,0}}{\rho_{m,0} \;
exp \left( -3 \int_{a_{0}}^{a} (1+\gamma(a'))\frac{da'}{a'} \right)}   
\left( \frac{a}{a_{0}} \right)^{-3(1+w(a))-1}  \nonumber \\
& \times &\left[ \frac{a}{a_{0}} \ln \left(\frac{a}{a_{0}} \right) dw(a) +
(w(a)-\eta(a)) d\left( \frac{a}{a_{0}} \right) \right] \, .
\end{eqnarray}
Generally, it is not possible to solve this equation in closed form, so
further in this section we consider a simplified model with 
$\gamma(a)=\gamma=const$, 
$\eta(a)=\eta=const$ and $w(a)=w=const$, to gain deeper insight via an analytical
solution which one can obtain in this case. The function $G$ can now be
expressed in terms of $a$ as
\begin{equation}
\label{eq:GEOSconst}
G = G_{0} \left(1 - \frac{\rho_{d,0}}{\rho_{m,0}} \frac{\eta-w}{\gamma-w}
 \left[ \left( \frac{a}{a_{0}} \right)^{-3(w-\gamma)}-1 \right] \right)\, .
\end{equation}
Once we have the expression for $G$, we can give the expression for the
other source term in the Friedmann equation (the first is given by 
(\ref{eq:Glaw})):
\begin{equation}
\label{eq:Grhom}
G \rho_{m} = G_{0} \left( \rho_{m,0}+\rho_{d,0} \frac{\eta-w}{\gamma-w} 
\right) \left( \frac{a}{a_{0}} \right)^{-3(1+\gamma)} -
G_{0}\rho_{d,0} \frac{\eta-w}{\gamma-w} 
\left( \frac{a}{a_{0}} \right)^{-3(1+w)} \, ,
\end{equation}
while the total source term (the right-hand side of the 
Friedmann equation) becomes
\begin{equation}
\label{eq:Grhototal}
G (\rho_{m}+ \rho_{d})= G_{0} \left( \rho_{m,0}+\rho_{d,0} \frac{\eta-w}{\gamma-w} 
\right) \left( \frac{a}{a_{0}} \right)^{-3(1+\gamma)} +
G_{0}\rho_{d,0} \frac{\gamma-\eta}{\gamma-w} 
\left( \frac{a}{a_{0}} \right)^{-3(1+w)} \, .
\end{equation}
Closer inspection of equation (\ref{eq:Grhototal}) shows that in
our cosmological model the universe evolves as if the quantity $G$ 
were constant and we had one
phantom component with the parameter of EOS $w$ and one non-phantom component
with the parameter of EOS $\gamma$, both components having conserved
energy-momentum tensors. Therefore, our cosmological model mimics the behaviour
of the model with a non-phantom matter component with the present energy density
$\tilde{\rho}_{m,0} = \rho_{m,0}+\rho_{d,0} \frac{\eta-w}{\gamma-w}$ and a
phantom component with the present energy density
$\tilde{\rho}_{d,0} = \rho_{d,0} \frac{\gamma-\eta}{\gamma-w}$. The acceleration
of the universe in our model is given by
\begin{eqnarray}
\label{eq:acc}
\frac{\ddot{a}}{a} & = & -\frac{4 \pi}{3} G_{0} \left[ (1+3\gamma) 
\left( \rho_{m,0}+\rho_{d,0} \frac{\eta-w}{\gamma-w} 
\right) \left( \frac{a}{a_{0}} \right)^{-3(1+\gamma)} \right. \nonumber \\
& + &
\left. \left(1+3\eta-(1+3\gamma)\frac{\eta-w}{\gamma-w} \right)
\rho_{d,0} \left( \frac{a}{a_{0}} \right)^{-3(1+w)} \right] \, .
\end{eqnarray}

Let us finally consider possible constraints on the parameters of the model.
The assumption of constancy of all parameters of EOS is probably an
oversimplification. However, for modest variations of some of the parameters 
with
the scale factor $a$, one expects that the model given above represents a good
approximation. It is certainly conceivable that for the more realistic parameters
$w(a)$, $\eta(a)$ and $\gamma(a)$ one can, at the level of the Friedmann
equation, obtain in our model a sort of dynamics which is 
identical with the dynamics in
the more general model with constant $G$ and conserved energy-momentum tensors 
of the phantom and
non-phantom components. Given many observational constraints on the past
evolution of $G$ \cite{Gconstr}, one would expect quite stringent constraints on the
difference $\eta-w$, which should be small. This means that this model could
explain the cosmological expansion with $w$ not much more negative than $-1$. 
However, in the general case of our model,
a sort of cosmology where $w$ would differ from $-1$ more
substantially is certainly not excluded.    
The expression (\ref{eq:acc}) for the acceleration of the expansion of the
universe provides another constraint on the parameters of the model. Namely, in
order to have a transition from deceleration to acceleration at low
redshift ($z \sim 1$), the coefficient $1+3\eta-(1+3\gamma)\frac{\eta-w}
{\gamma-w}$ must be negative. In the case that the difference $\eta-w$ is small,
this requirement reduces to the standard one, $\eta < -1/3$. The requirement of
$\eta-w$ being small also favours the possibility $\eta=-1$,
which is equivalent to the time-dependent cosmological constant. Models with the
time-dependent cosmological constant and $G$, studied in (\cite{Gvar}), represent
a specially interesting case. Models with a growing cosmological constant (and
a time-dependent $G$) \cite{lamgrow} exhibit a very peculiar fate of the universe,
leading to the unbounding of all gravitationally bound systems, while leaving
non-gravitationally bound systems unaffected, the so-called   
``partial rip" scenario. 

\section{A model with two interacting components}

\label{sec3}

In this section we consider a model with two interacting, non-phantom
components. The first component, the dark energy component, is described by the
equation of state
\begin{equation}
\label{eq:darkint}
p_{d}=\eta(a) \rho_{d}\, , \;\;\;\;\;\; \eta \ge -1 \, ,
\end{equation}
while the other component, the matter component, is determined by the following
equation of state: 
\begin{equation}
\label{eq:matint}
p_{m}=\gamma(a) \rho_{m} \, , \;\;\;\;\;\; \gamma \ge 0 \, .
\end{equation}
In this model, the gravitational constant $G$ has no space-time variation.
The interaction of the components is included in the model in the following way.
We assume that the energy-momentum tensors of 
the two separate components are not conserved,
but the total energy-momentum tensor $T^{\mu \nu} =  T^{\mu \nu}_{m}+T^{\mu
\nu}_{d}$ is conserved. In this way, there exists an exchange of energy and
momentum between the two components. The requirement of the conservation of the
total energy-momentum tensor can be expressed as
\begin{equation}
\label{eq:totint}
d\rho_{m} +3\rho_{m}(1+\gamma(a))\frac{da}{a} = 
-d\rho_{d} -3\rho_{d}(1+\eta(a))\frac{da}{a} \, .
\end{equation}
What remains to be determined is the specification of the interaction
(energy-momentum exchange) between the components. The aim of this model is to
demonstrate that this set-up can mimic the expansion of the universe
characteristic of phantom cosmologies. Therefore we assume the following
evolution law for the dark energy component:  
\begin{equation}
\label{eq:rhodint}
\rho_{d} = \rho_{d,0} \left( \frac{a}{a_{0}} \right)^{-3(1+w(a))} \, ,
\end{equation}
where $w(a)< -1$, i.e. the dark energy component has the evolution law
characteristic of phantom energy. The non-phantom dark energy component has the
scaling with $a$ characteristic of phantom energy owing to the interaction with
the matter component.
Equation (\ref{eq:totint}) then determines
the evolution law for the energy density of the matter component. 
For general values of the
parameters of EOS it is not always possible to obtain the solutions in closed
form. Therefore, in the remainder of this section we assume that these
parameters are constant, i.e. $\gamma(a)=\gamma=const$, $\eta(a)=\eta=const$ and
$w(a)=w=const$. This particular choice will allow us to gain insight via closed
form solutions. The energy density of the matter component then becomes  
\begin{equation}
\label{eq:rhomint}
\rho_{m} = \left(\rho_{m,0} + \rho_{d,0} \frac{\eta-w}{\gamma-w} \right)
\left( \frac{a}{a_{0}} \right)^{-3(1+\gamma)} -
\rho_{d,0} \frac{\eta-w}{\gamma-w}
\left( \frac{a}{a_{0}} \right)^{-3(1+w)}\, .
\end{equation}
The total energy density, $\rho=\rho_{m}+\rho_{d}$, which appears on the
right-hand side of equation (\ref{eq:Friedmann}), then has the form
\begin{equation}
\label{eq:rhototint}
\rho = \left(\rho_{m,0} + \rho_{d,0} \frac{\eta-w}{\gamma-w} \right)
\left( \frac{a}{a_{0}} \right)^{-3(1+\gamma)} +
\frac{\gamma-\eta}{\gamma - w} \rho_{d,0}
\left( \frac{a}{a_{0}} \right)^{-3(1+w)}\, .
\end{equation}
The acceleration of the expansion of the universe is given by the expression
\begin{eqnarray}
\label{eq:accint}
\frac{\ddot{a}}{a}& = & -\frac{4 \pi}{3} G  \left[
(1+3\gamma) \left(\rho_{m,0} + \rho_{d,0} \frac{\eta-w}{\gamma-w} \right)
\left( \frac{a}{a_{0}} \right)^{-3(1+\gamma)}  \right. \nonumber \\
& + &
 \left. \left(1+ 3\eta -(1+3\gamma) \frac{\eta-w}{\gamma-w} \right) \rho_{d,0}
\left( \frac{a}{a_{0}} \right)^{-3(1+w)} \right] \, .
\end{eqnarray}
As for the model displayed in section \ref{sec2}, the right-hand side
of the Friedmann equation is the same as in a model with constant $G$ and two
non-interacting components: the first being the non-phantom component with 
the present energy density $\tilde{\rho}_{m,0} = 
\rho_{m,0} + \rho_{d,0} \frac{\eta-w}{\gamma-w}$
and the parameter of EOS $\gamma$, while the second being phantom energy with
the present energy density $\tilde{\rho}_{d,0} = 
\frac{\gamma-\eta}{\gamma-w} \rho_{d,0}$
and the parameter of EOS $w$. Equation (\ref{eq:rhomint}) shows the
effects of the interaction with the dark energy component on
$\rho_{m}$ as an additional term
growing as $a^{-3(1+w)}$. The requirement that the scaling law of the matter
component should not differ too much from the scaling law dictated by its EOS 
($\sim a^{-3(1+\gamma)}$), i.e. that the interaction is not too strong, leads to the
condition that the difference $\eta-w$ should be small. In the model with a more
general variation of some of the parameters $\gamma$, $\eta$ or $w$, it is 
conceivable that this constraint would be milder. Again, as in section 
\ref{sec2}, two non-phantom components mimic phantom cosmology. The model can
successfully describe the transition from the decelerating to the accelerating
regime of the expansion of the universe if the coefficient 
$1+ 3\eta -(1+3\gamma) \frac{\eta-w}{\gamma-w}$ is negative. When $\eta$ is
close to $w$, the afore-mentioned requirement reduces to the condition 
$\eta < -1/3$. One especially interesting variant of the model is the case
$\eta = -1$. In this case, the evolving cosmological constant in interaction with
the matter component mimics the expansion of phantom cosmology.

One way of elaborating the model given in this section would certainly be its
formulation in terms of classical fields. We can model the system of two
interacting components as a system of two minimally coupled interacting scalar 
fields in a cosmological setting. For the Lagrangian of the interacting system we
then take a general form (we consider only time-dependent scalar fields)
\begin{equation}
\label{eq:L}
{\cal L} = \frac{\dot{\phi}^{2}}{2} + \frac{\dot{\psi}^{2}}{2} - V(\phi,\psi)
\, ,
\end{equation}
where $\phi$ and $\psi$ denote scalar fields. Given that the total energy
density is $\rho = \frac{\dot{\phi}^{2}}{2} + \frac{\dot{\psi}^{2}}{2} 
+ V(\phi,\psi)$ and the total pressure is $p = {\cal L}$, one obtains the
following two constraints on the dynamics of the scalar fields:
\begin{eqnarray}
\label{eq:scalar}
\dot{\phi}^{2} + \dot{\psi}^{2} & = & (1+\eta)\rho_{d} + (1+\gamma)\rho_{m} 
\, , \nonumber  \\
2 V(\phi, \psi) & = & (1-\eta)\rho_{d} + (1-\gamma)\rho_{m} \, .
\end{eqnarray}
From (\ref{eq:rhodint}) and (\ref{eq:rhomint}) we have obtained $\rho_{d}$ and
$\rho_{m}$, respectively, as functions of the scale factor $a$. On the other
hand, from (\ref{eq:Friedmann}) we can determine the function $a(t)$. This
makes the right-hand sides of equations (\ref{eq:scalar}) known functions of
time. All pairs of the functions $\phi$ and $\psi$  (with a nontrivial potential
$V(\phi, \psi)$) that satisfy equations 
(\ref{eq:scalar}) can produce the evolution of the universe as described in the
model of this section. This class of solutions certainly does not
exclude more sophisticated (and realistic) field (or microscopic) models.

\section{Conclusions}

\label{sec4}

The two models, described in sections \ref{sec2} and \ref{sec3}, 
have been constructed to
demonstrate that cosmologies {\em without} phantom energy can lead to an expansion of
the universe usually attributed to phantom energy. The first model is
characterized by an evolving gravitational ``constant" $G$ and a dark energy
component with a non-conserved energy-momentum tensor. The second model is based
on two interacting components. Both models yield results for the cosmological
evolution of their components which are testable against the results of various
cosmological observations. Calculations in this paper have been made with a specific
choice of parameters (e.g. constant parameters of EOS) which ensures closed form
solutions. These solutions facilitate the interpretation of the physical meaning
of the obtained results, but the scope of 
the models described in this paper certainly does not
end here. Models with variable (e.g. dependent on $a$) parameters $\gamma$,
$\eta$ and $w$ offer much more possibilities (especially in terms of 
satisfying numerous
constraints from the past evolution of the universe) and merit further
investigation. The possibility of mimicking phantom cosmology by non-phantom
one, certainly does not rule out an appealing and provocative idea of phantom
energy. However, it puts a greater ponder on the nature of phantom
energy as an effective description of the more complex dynamics of non-phantom
matter.

{\bf Acknowledgements.} The author would like to thank B. Guberina, N. Bili\'{c}
and R. Horvat for useful comments on the manuscript. This work was
supported by the Ministry of Science, Education and Sport of the Republic of 
Croatia under the contract No. 0098002.


\begin{thebibliography}{88}
\bibitem{SN}  A.G. Riess et al., Astron. J. 116 (1998) 1009; S. Perlmutter et 
al., Astrophys. J. 517 (1999) 565.
\bibitem{CMBR} P. de Bernardis et al., Nature 404 (2000) 955; A.D. Miller et al.
Astrophys. J. Lett. 524 (1999) L1; S. Hanany et al., Astrophys. J. Lett. 545
(2000) L5; N.W. Halverson et al., Astrophys. J. 568 (2002) 38; B.S. Mason et
al., astro-ph/0205384; D.N. Spergel et al., astro-ph/0302209; L. Page et al.
astro-ph/0302220.
\bibitem{LSS}  R. Scranton et al., astro-ph/0307335; M. Tegmark et al., 
astro-ph/0310723.
\bibitem{Rev}  W.L. Freedman, M.S. Turner, astro-ph/0308418;
 S.M. Carroll, astro-ph/0310342.
\bibitem{singh} R.G. Vishwakarma, P. Singh, Class. Quant. Grav. 20 (2003) 2033;
V. Sahni, Y. Shtanov, JCAP 0311 (2003) 014. 
\bibitem{wein} S. Weinberg, Rev. Mod. Phys. 61 (1989) 1.
\bibitem{peeb} P.J.E. Peebles, B. Ratra, Rev. Mod. Phys. 75 (2003) 559.
\bibitem{pad} T. Padmanabhan, Phys. Rept. 380 (2003) 235.
\bibitem{mi}  A. Babic, B. Guberina, R. Horvat, H. Stefancic, 
Phys. Rev. D 65 (2002) 085002; B. Guberina, R. Horvat, H. Stefancic,
 Phys. Rev. D 67 (2003) 083001.
\bibitem{sola} I.L. Shapiro, J. Sola, Phys. Lett. B 475 (2000) 236; I.L. Shapiro, 
J. Sola, JHEP 0202 (2002) 006; I.L. Shapiro, J. Sola, C. Espana-Bonet, 
P. Ruiz-Lapuente, Phys. Lett. B 574 (2003) 149.
\bibitem{bonanno} A. Bonanno, M. Reuter, Phys. Lett. B 527 (2002) 9; 
E. Bentivegna, A. Bonanno, M. Reuter, astro-ph/0303150.
\bibitem{Q} B. Ratra, P.J.E. Peebles, Phys. Rev. D 37 (1988) 3406; 
P.J.E. Peebles, B. Ratra, Astrophys. J. 325 (1988) L17; C. Wetterich, Nucl.
Phys. B 302 (1988) 668; R.R. Caldwell, R. Dave, P.J. Steinhardt, Phys. Rev.
Lett. 80 (1998) 1582; I. Zlatev, L. Wang, P.J. Steinhardt, Phys Rev. Lett. 82
(1999) 896.
\bibitem{tach} A. Sen, JHEP 0204 (2002) 048; A. Sen, JHEP 0207 (2002) 065;
A. Sen, Mod. Phys. Lett. A 17 (2002) 1797;
T. Padmanabhan, T. Roy Choudhury, Phys. Rev. D 66 (2002) 081301;
J.S.Bagla, H.K.Jassal, T.Padmanabhan, Phys. Rev. D 67 (2003) 063504.
\bibitem{CG}  A. Yu. Kamenshchik, U. Moschella,  V. Pasquier, Phys. Lett. B511 
(2001) 265;  N. Bilic, G.B. Tupper, R.D. Viollier, Phys. Lett. B 535 (2002) 17;
 M.C. Bento, O. Bertolami, A.A. Sen, Phys. Rev. D 66 (2002) 043507. 
\bibitem{odman} A. Melchiorri, L. Mersini, C.J. Odman, M. Trodden, Phys.
Rev. D 68 (2003) 043509.
\bibitem{padman1} T. Padmanabhan, T. Roy Choudhury, Mon. Not. Roy. Astron. Soc. 
344 (2003) 823.
\bibitem{padman2} T. Roy Choudhury, T. Padmanabhan, astro-ph/0311622.
\bibitem{lima} J.A.S. Lima, J.V. Cunha, J.S. Alcaniz, Phys. Rev. D 68 
(2003) 023510; J.S. Alcaniz, astro-ph/0312424.
\bibitem{sahni}  U. Alam, V. Sahni, T. D. Saini, A. A. Starobinsky, 
astro-ph/0311364.
\bibitem{cald} R.R. Caldwell, Phys. Lett. B 545 (2002) 23.
\bibitem{tech} C. Armendariz-Picon, T. Damour, V. Mukhanov, Phys. Lett.
 B 458 (1999) 209; T. Chiba, T. Okabe, M. Yamaguchi, Phys. Rev. D 62 (2000) 
 023511; V. Faraoni, Int. J. Mod. Phys. D 11 (2002) 471; B. McInnes, 
 JHEP 0208 (2002) 029; S. Nojiri, S.D.
 Odintsov, Phys. Lett. B 562 (2003) 147; S. Nojiri, S.D. Odintsov, Phys. Lett 
 B 565 (2003) 1; P. Singh, M. Sami, N. Dadhich, Phys. Rev. D 68 (2003) 023522;
 G.W. Gibbons, hep-th/0302199; L.P.Chimento, R. Lazkoz, gr-qc/0307111; J.G. Hao,
 X.Z. Li, Phys. Rev. D 68 (2003) 083514; Y-S. Piao, E. Zhou, Phys. Rev. D 68 
 (2003) 083515; M.P.Dabrowski, T. Stachowiak, M.
 Szydlowski, hep-th/0307128; J.G. Hao, X.Z. Li Phys. Rev. D 67 (2003) 107303;
 J.G. Hao, X.Z. Li, astro-ph/0309746; V. Faraoni, gr-qc/0307086; V.B. Johri, 
 astro-ph/0311293;  M. Sami, A. Toporensky, gr-qc/0312009;
 E. Elizalde, J. Quiroga Hurtado, Mod. Phys. Lett. A19 (2004) 29;
 E. Elizalde, S. Nojiri, S.D. Odintsov, hep-th/0405034. 
\bibitem{cald2}  R.R. Caldwell, M. Kamionkowski, N.N. Weinberg, Phys. Rev. Lett. 
91 (2003) 071301.
\bibitem{dec1} S.M. Carroll, M. Hoffman, M. Trodden, Phys. Rev. D 68 (2003) 
023509.
\bibitem{dec2} J.M. Cline, S. Jeon, G.D. Moore, hep-ph/0311312.
\bibitem{GPE} H. Stefancic, Phys. Lett. B 586 (2004) 5, astro-ph/0310904.
\bibitem{odin} E. Elizalde, S.D. Odintsov, Phys. Lett. B 303 (1993) 240; E.
Elizalde, S.D. Odintsov, Phys. Lett. B 321 (1994) 199; E. Elizalde, S.D.
Odintsov, I.L. Shapiro, Class. Quant. Grav. 11 (1994) 1607; E. Elizalde, C.
Lousto, S.D. Odintsov, A. Romeo, Phys. Rev. D 52 (1995) 2202.
\bibitem{extra} P. Loren-Aguilar, E. Garcia-Berro, J. Isern, Yu.A. Kubyshin, 
Class. Quant. Grav. 20 (2003) 3885.
\bibitem{scalten} see e.g. R. Nagata, T. Chiba, N. Sugiyama, astro-ph/0311274.
\bibitem{Gconstr} see e.g. C.J. Copi, A.N. Davis, L.M. Krauss, astro-ph/0311334.
\bibitem{Gvar} O. Bertolami, Nuovo Cim. B 93 (1986) 36;
A. Beesham, Nuovo Cimento 96 B (1986) 17; A. Beesham, Int. J.
Theor. Phys. 25 (1986) 1295; A-M.M. Abdel-Rahman, Gen. Rel. Grav. 22 (1990) 655;  
M.S. Berman, Phys. Rev. D 43 (1991) 1075; M.S.
Berman, Gen. Rel. Grav. 23 (1991) 465; R.F. Sistero, Gen. Rel. Grav. 32 (1991)
1265; D. Kalligas, P. Wesson, C.W.F. Everitt, Gen. Rel. Grav. 24 (1992) 351; 
T. Singh, A. Beesham, Gen. Rel. Grav. 32 (2000) 607; A.I. Arbab, 
A. Beesham, Gen. Rel. Grav. 32 (2000) 615; A.I. Arbab, Spacetime and Substance 
1(6) (2001) 39; A.I. Arbab, astro-ph/0308068; J. Ponce de Leon, gr-qc/0305041.
\bibitem{lamgrow} H. Stefancic, to appear in Phys. Lett. B, astro-ph/0311247.

\end{thebibliography}
\end{document}